\documentclass[preprint,aps,amsmath,pre,showpacs]{revtex4-1}
\usepackage{graphicx}
\usepackage{amsfonts}
\usepackage[english]{babel}

\begin{document}

\title{Speciation in the Derrida-Higgs model with finite genomes and spatial populations}

\author{Marcus A.M. de Aguiar}

\affiliation{Instituto de F\'{i}sica `Gleb Wataghin',
Universidade Estadual de Campinas, Unicamp\\ 13083-970, Campinas,
SP, Brazil}

\begin{abstract}

The speciation model proposed by Derrida and Higgs demonstrated that a sexually reproducing
population can split into different species in the absence of natural selection or any type of
geographic isolation, provided that mating is assortative and the number of genes involved in the
process is infinite. Here we revisit this model and simulate it for finite genomes, focusing on the
question of how many genes it actually takes to trigger neutral sympatric speciation. We find that,
for typical parameters used in the original model, it takes of the order of $10^5$ genes. We compare
the results with a similar spatially explicit model where about 100 genes suffice for speciation. We
show that when the number of genes is small the species that emerge are strongly segregated in
space. For larger number of genes, on the other hand, the spatial structure of the population is
less important  and the species distribution overlap considerably.

\end{abstract}

\maketitle

\section{Introduction}
\label{intro}

Speciation requires the evolution of reproductive isolation between initially compatible individuals
\cite{mayr_1942}. One of the most accepted mechanisms for the development of reproductive barriers
is the geographic isolation of the populations, which can be complete (allopatry) or partial
(parapatry) \cite{coyne_speciation_2004,nostil_2012}. In both cases the reduction of gene flow
between individuals inhabiting different regions facilitates the fixation of local adaptations.
These, in turn, may lead to pre-zygotic or post-zygotic mating incompatibilities and can eventually
develop into full reproductive isolation.

Sympatric speciation, where new species arise from a population inhabiting a single geographic
region, has been highly debated \cite{kirk-2004,Bolnick_2007} and its occurrence in nature is
documented in only a few cases \cite{kondra-1986,coyne_speciation_2004}. Mathematical models have
shown that sympatric speciation is theoretically possible
\cite{Udovic_1980,Felsenstein_1981,kondra-1998,kondra-1999,Doebeli_2005,Gavrilets_2006} and the
model by  Dieckmann and Doebeli \cite{Dieckmann_1999}, in particular, attracted a lot of attention.
In their model, speciation results from the interplay between strong competition for resources and
the way resources are distributed. It was shown that, under appropriate conditions, the population
will evolve in such a way as to consume the most abundant type of resource, but it will arrive at
the corresponding `optimal' phenotype in a minimum of the fitness landscape, causing it to split in
two. The resulting system of two subpopulations with different phenotypes would then be at a fitness
maximum. For sexually reproducing individuals the evolution of reproductive barriers between the two
nearly split populations would need a further ingredient, assortative mating. This is tendency of
individuals to mate with others that are similar to themselves and prevents the mixing of the two
subpopulations. In the model assortative mating was shown to evolve naturally because only then the
fitness maximum could be achieved. In spite of its theoretical plausibility, this and other models
have been criticized for their supposedly unrealistic assumptions
\cite{gavrilets_fitness_2004,Barton_2005,Bolnick_2004,Gavrilets_2005}.

More surprising than sympatric speciation under strong selection is the possibility of sympatric
speciation in a neutral scenario \cite{kondra-1998}. It was shown by Derrida and Higgs (DH)
\cite{higgs_stochastic_1991,higgs_genetic_1992} that a population of initially identical individuals
subjected only to mutations and assortative mating  can indeed split into species in sympatry if the
genomes are infinitely large. The hypothesis of genomes containing infinitely many loci, though not
new \cite{Kimura_1964,Ewens_1979}, raises important questions about the meaning of a locus (a gene,
the position of a base-pair on a chromosome or other replicating molecule) and the actual number of
such loci that would be necessary for speciation to occur, since, after all, genes or base-pairs are
always finite. Moreover, the model assumes that all these infinitely many loci are involved in
assortative mating, which also needs to be taken into account in the interpretation of the loci.

In this paper we consider the DH model for finite genomes. We find that, for the parameters
considered in the original paper, speciation occurs only for very large number of loci, of the order
of $10^5$. We compare the dynamics of the model for finite number of loci with the infinite loci
model and show how the broadening of the genetic distribution of genotypes in the finite loci model
prevents the onset of speciation. We also compare the finite loci DH model with a spatial version of
the dynamics introduced in \cite{de_aguiar_global_2009}. When space is added and reproduction is
restricted by assortative mating and spatial proximity, the minimum number of loci needed for
speciation drops by several orders of magnitude and can occur even for as few as 100 loci. Our
simulations show that when the number of loci is small species form in well defined spatial regions,
with little overlap at the boundaries. When the number of loci is large, on the other hand, space
becomes less important and the species overlap considerably more in space.

\section{The Derrida-Higgs model of sympatric speciation}
\label{DH}

The model introduced by  Derrida and Higgs \cite{higgs_stochastic_1991} considers a sympatric
population of $M$ haploid individuals whose genomes are represented by binary strings of size $B$,
$\{S_1^\alpha, S_2^\alpha, \dots, S_B^\alpha\}$ where $S_i^\alpha$, can assume the values $\pm 1$.
For simplicity, each locus of the genome will be called a gene and the values $+1$ and $-1$ the
corresponding alleles. The number of individuals at each generation is kept constant and the
population is characterized by an $M \times M$ matrix $q$ measuring the degree of genetic similarity
between pairs of individuals: 
\begin{equation}
q^{\alpha \beta} = \frac{1}{B} \sum_{i=1}^B S_i^\alpha S_i^\beta.
\label{qdef}
\end{equation}
If the genomes of $\alpha$ and $\beta$ are identical $q^{\alpha \beta}=1$ whereas two genomes with
random entries will have $q^{\alpha \beta}$ close to zero. Alternatively, the genetic distance
between the individuals, measuring the number of genes bearing different alleles, is $d^{\alpha
	\beta} = B(1 - q^{\alpha \beta})/2$.

Each generation is constructed from the previous one as follows: a first parent $P_1$ is chosen at
random. The second parent $P_2$ has to be genetically compatible with the first, i.e., their degree of
similarity has to satisfy $q^{P_1 P_2} \geq q_{min}$. In other words parents must have at least $G =
B (1-q_{min})/2$ genes bearing the same allele to be compatible (assortative mating). Individuals
$P_2$ are then randomly selected until this condition is met. If no such individual is found, $P_1$
is discarded and a new first parent is selected. The offspring inherits, gene by gene, the allele of
either parent with equal probability (sexual reproduction). The process is repeated until $M$
offspring have been generated. Individuals are also subjected to a mutation rate $\mu$ per gene,
which is typically small. 

The model is neutral in the sense that the probability that an individual is chosen as first parent
($1/M$) does not depend on its genome. However, once the first parent has been selected, the
chances of being picked as second parent and produce an offspring does depend on the genome and a
fitness measure can actually be defined in association with assortativeness \cite{de_aguiar_error_2015}.

To understand how the similarity matrix changes through generations, consider first an asexual
population where each individual $\alpha$ has a single parent $P(\alpha)$ in previous generation.
The allele $S_i^\alpha$ will be equal to $S_i^{P(\alpha)}$ with probability
$\frac{1}{2}(1+e^{-2\mu}) \approx 1 -\mu$ and $-S_i^{P(\alpha)}$ with probability
$\frac{1}{2}(1-e^{-2\mu}) \approx \mu$, so that the expected value is 
\begin{equation}
E(S_i^\alpha) = e^{-2\mu} S_i^{P(\alpha)}.
\label{aves}
\end{equation}
For independent genes, the expected value of the similarity between $\alpha$ and $\beta$ is, therefore,
 \begin{equation}
 E(q^{\alpha \beta}) = e^{-4\mu} q^{P(\alpha) P(\beta)}.
 \label{aveq}
 \end{equation}

In sexual populations $\alpha$ and $\beta$ have two parents each, $P_1(\alpha)$, $P_2(\alpha)$ and
$P_1(\beta)$,  $P_2(\beta)$, respectively, and since each inherits (on the average) half the alleles
from each parent, it follows that, on the average,
\begin{equation}
q^{\alpha \beta} = \frac{e^{-4\mu}}{4} \left( q^{P_1(\alpha) P_1(\beta)} +  q^{P_2(\alpha) P_1(\beta)} 
+ q^{P_1(\alpha) P_2(\beta)}  +   q^{P_2(\alpha) P_2(\beta)}  \right)
\label{hdupdate}
\end{equation}
with $q^{\alpha \alpha} \equiv 1$.  

In the limit of infinitely many genes, $B \rightarrow \infty$, this expression becomes exact and the
entire dynamics can be obtained by simply updating the similarity matrix. If there is no restriction
on mating, $q_{min}=0$, the overlaps $q^{\alpha \beta}$ converge to a stationary  distribution
centered at $q_0 \approx 1/(1+4\mu M)$. The approximation holds for $\mu$ and $1/M $ much smaller
than one, which is always the case for real populations. If $q_{min} > q_0$ the population splits
into species formed by groups of individuals whose average similarity is larger than $q_{min}$ and
such that interspecies similarity is smaller than $q_{min}$, tending to zero with time
\cite{higgs_stochastic_1991}.

\begin{figure}
	\centering
	\includegraphics[scale=0.25]{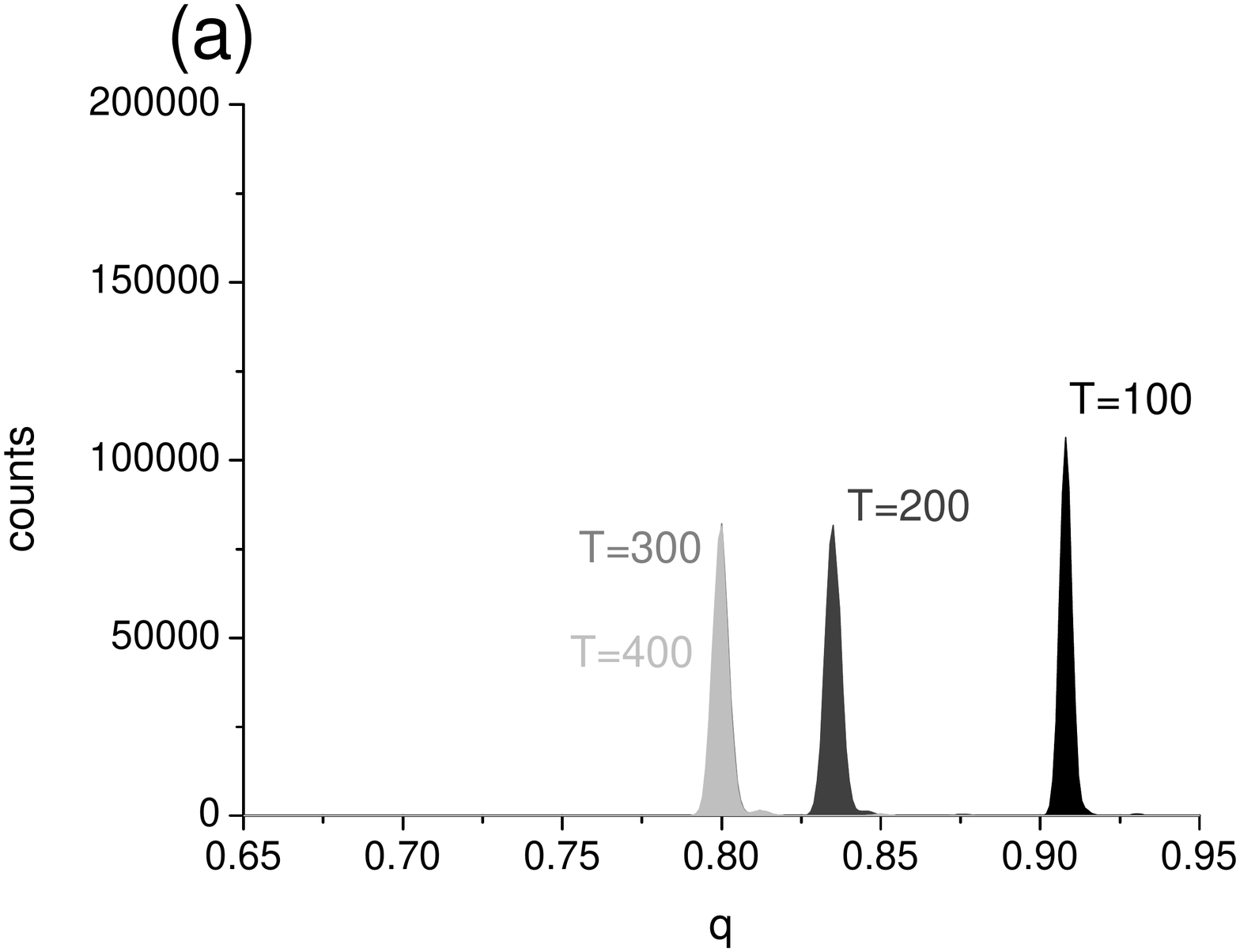}
	\includegraphics[scale=0.25]{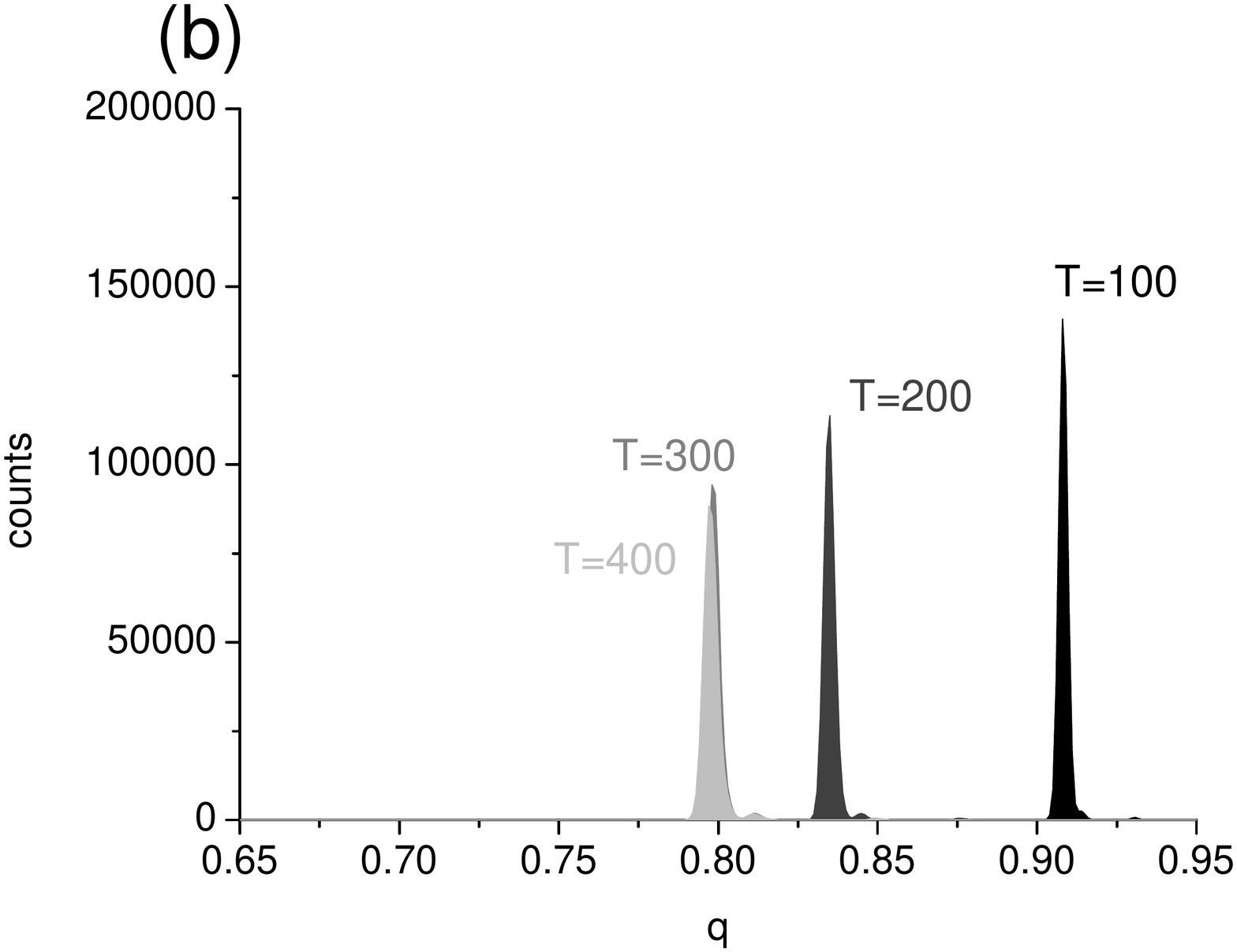}
	\includegraphics[scale=0.25]{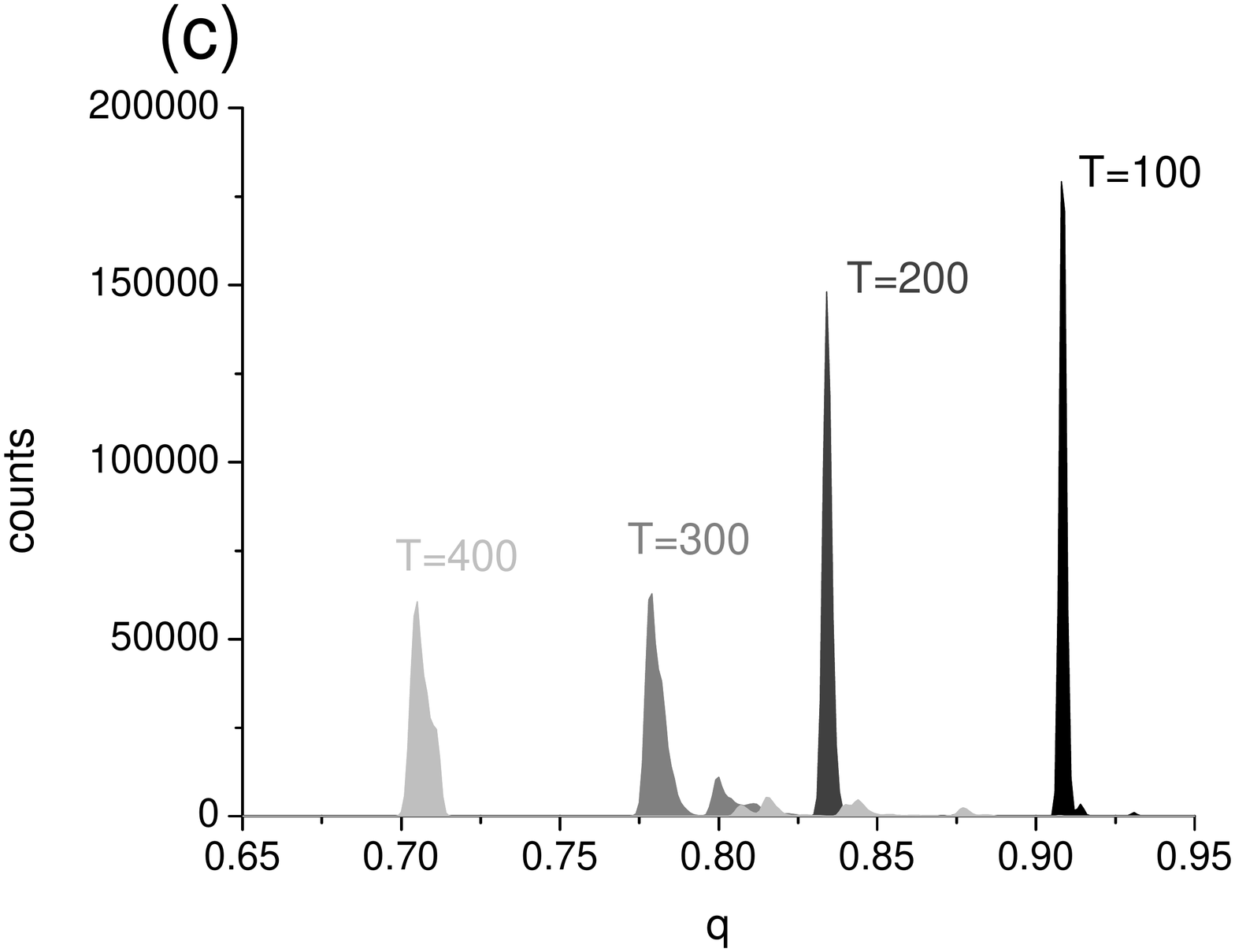}
	\includegraphics[scale=0.24]{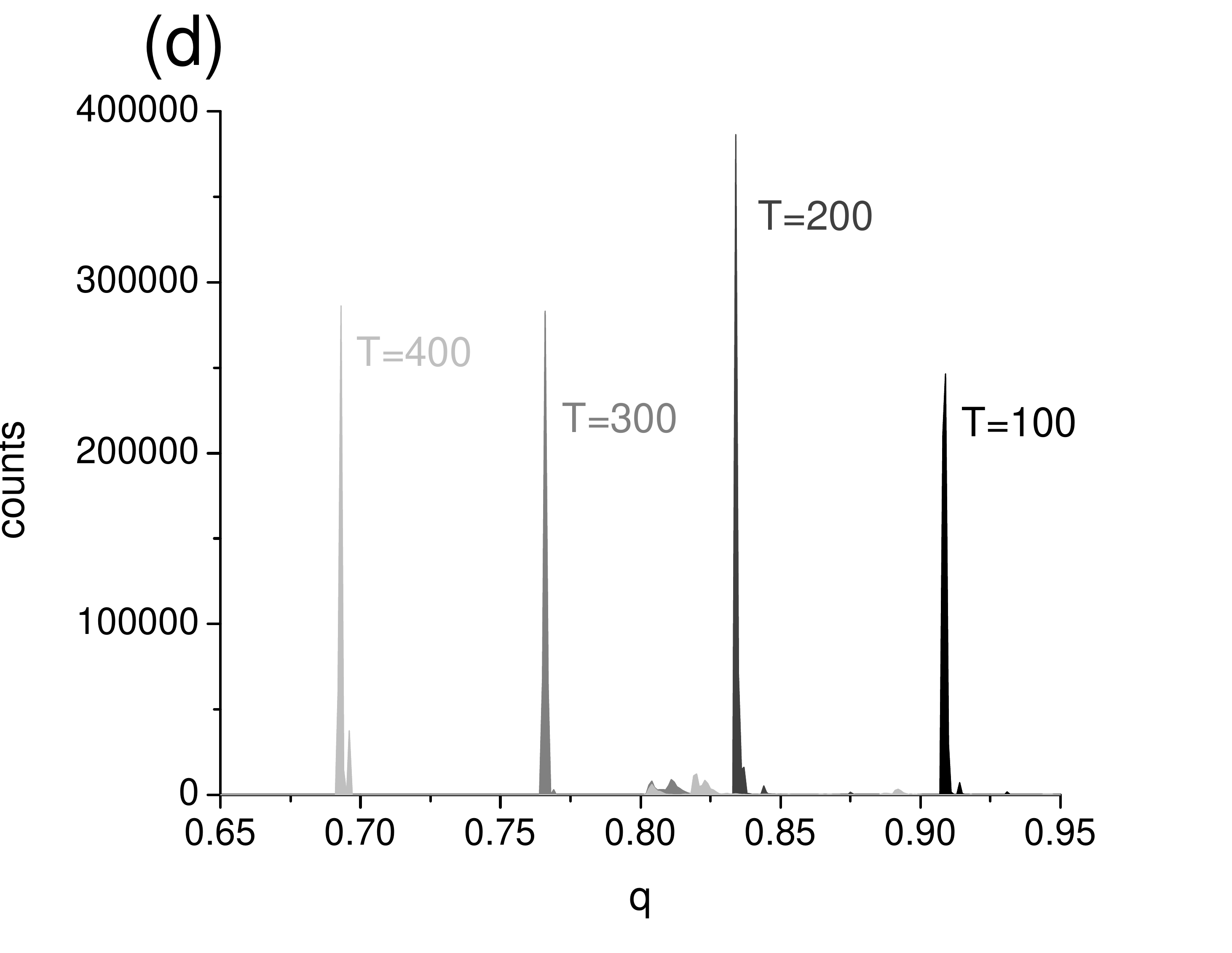}
	\caption{Distribution of similarity coefficients for the DH model with  (a) 50,000, (b) 100,000 and 
	(c) 200,000 genes. Panel (d) shows the result for original DH model with infinite genes. The distribution
	is shown after $T=100$, $200$, $300$ and $400$ generations. In all cases $M=1000$, $\mu=1/4000$ and 
    $q_{min}=0.8$.}
	\label{fighd}
\end{figure}

Figure \ref{fighd}(d) shows the histogram of $q^{\alpha \beta}$ between all pairs of individuals for
$M=1000$, $\mu=1/4000$ ($q_0=0.5$) and $q_{min}=0.8$ for 100, 200, 300 and 400 generations. The
initial condition is  $q^{\alpha \beta}=1$ for all $\alpha$ and $\beta$, representing genetically
identical individuals at the start of the simulation. At $T=100$ and $T=200$ the histogram displays
peaks to right of $q_{min}$ showing that all pairs are still compatible. At $T=300$, on the other
hand, the peak at $0.77 < q_{min}$ is a signature of speciation, showing that several pairs of
individuals have become reproductively isolated. The corresponding species are represented by the
smaller peaks to the right of $q_{min}$.

\begin{figure}
	\centering
	\includegraphics[scale=0.25]{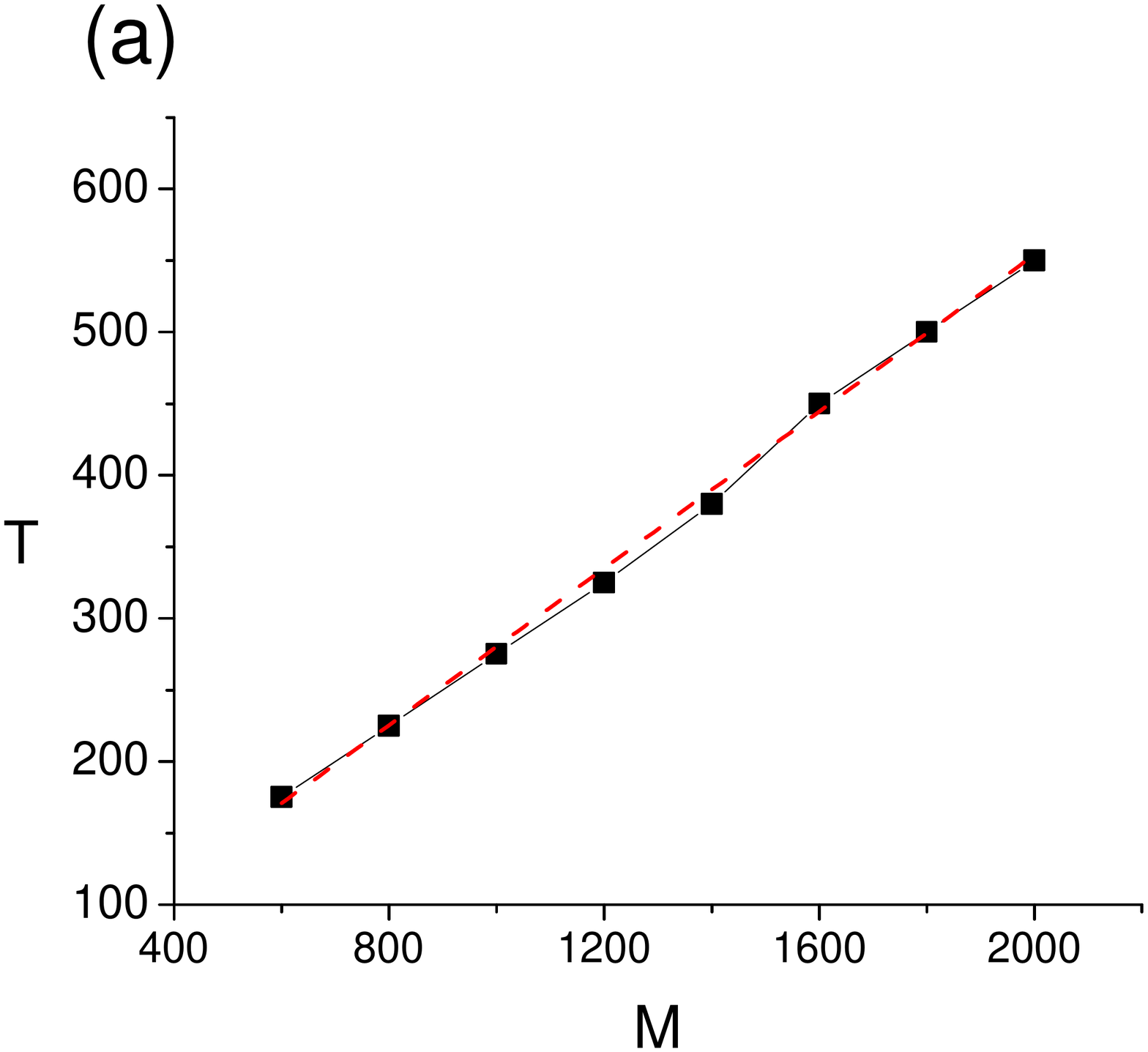}
	\includegraphics[scale=0.25]{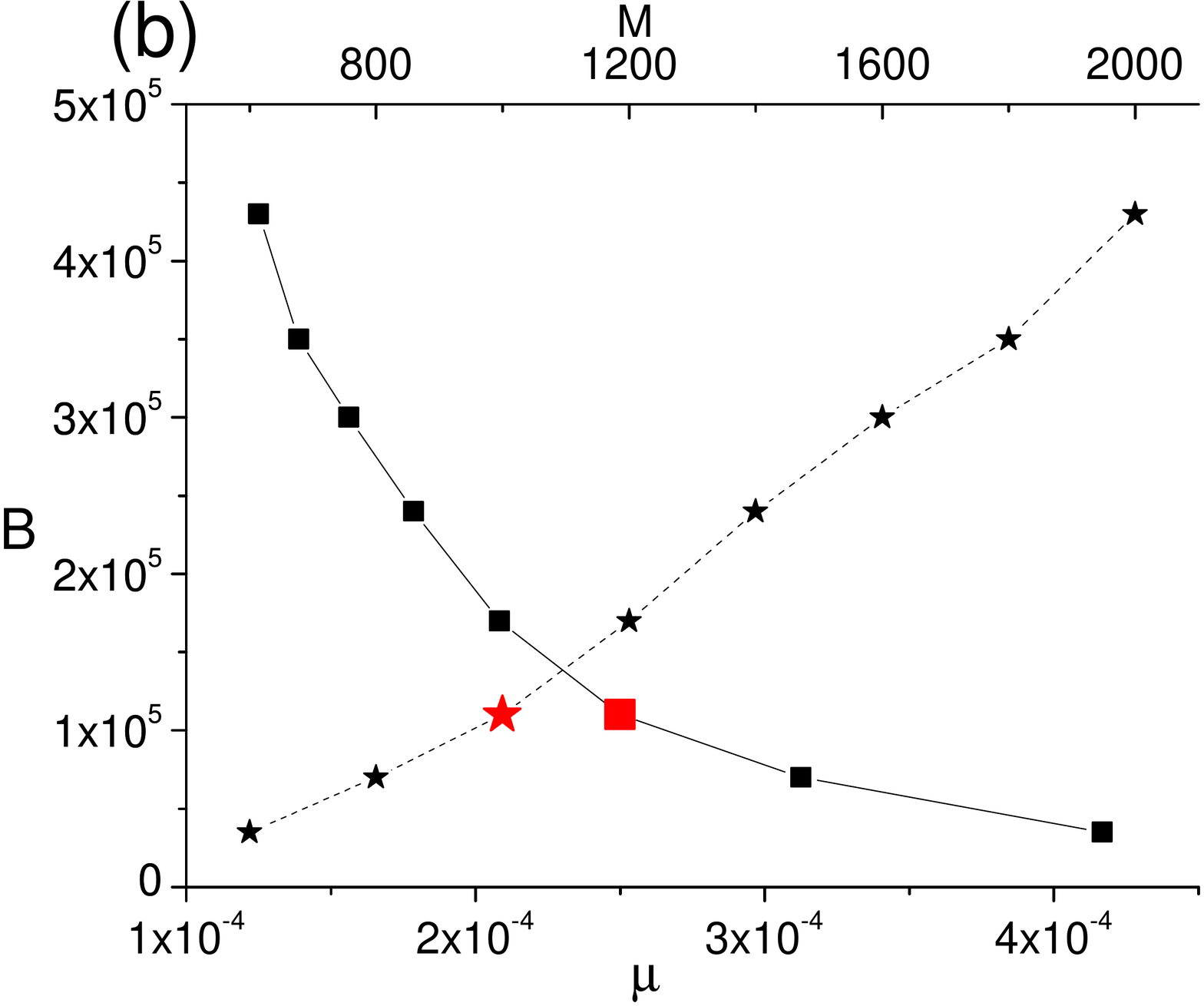}
	\caption{(color online) (a) Time to speciation as a function of population size  in the DH model. Dashed line
		(red) shows a linear fit. 
		(b)  Minimum genome size for speciation in the finite genome model as a function of $M$ (stars, upper
	horizontal axis) and $\mu$ (squares, lower horizontal axis). The values $M=1000$ and $\mu=0.00025$ used
	in the other figures are marked with a larger (red) symbol. In both panels $4\mu M=1$.}
	\label{time}
\end{figure}

The only condition for speciation in the DH model is $q_{mim} > q_0 \approx (1+4\mu M)^{-1}$
\cite{higgs_stochastic_1991}. Changing the mutation rate (or the population size) but keeping $\mu
M$ fixed only affects the time to speciation, which increases approximately linearly with $M$, or
$1/\mu$, as shown in Fig. \ref{time}(a). This is consistent with Eqs.(\ref{aveq}) and
(\ref{hdupdate}), which indicade that the change in the similarity matrix per time step is
proportional to $\mu$ for $\mu \ll 1$. Increasing $\mu M$, on the other hand, decreases $q_0$ and
increases the number of species formed \cite{higgs_stochastic_1991,higgs_genetic_1992}. 

\section{The DH model with finite genomes}
\label{DHF}

An important feature of the DH model is that there is no need to describe the genomes explicitly:
all it takes is the initial similarity matrix $q$, which is set to 1, and the update rule Eq.
(\ref{hdupdate}). To implement the model with a finite number of genes it is necessary to keep track
of all the $M$ genomes at each generation and calculate the similarities from the definition Eq.
(\ref{qdef}). Offspring for the next generation are created by choosing (gene by gene) the allele of
each parent with equal probability and letting the allele mutate (from +1 to -1 or vice-versa) with
probability $\mu$. Parents are chosen like in the original DH model, picking the first parent at
random and a second parent compatible with it.

If the number $B$ of genes is small the distribution of similarities starts from $q=1$ and moves to
the left until $q=q_{min}$, where it remains stationary and no speciation occurs. We found that
speciation only occurs for very large values of $B$, as shown in Fig. \ref{fighd} (a)-(c). For the
parameters used ($M=1000$, $\mu=0.00025$) we observed speciation only for $B$ larger than about
$110000$. When compared to the DH model we see that the finite number of genes blurs the peaks of
the $q$ distribution, preventing them from breaking up.

The minimum number of genes required for speciation depends directly on the mutation rate and
population size. Figure \ref{time}(b) shows the minimum value of $B$ for speciation as a function of
$M$ and $\mu$ for $4 \mu M = 1$. For populations with only $M=600$ individuals $B$ drops to about
35000 whereas for $M=2000$ it is necessary at least 430000 genes, showing that $B$ rapidly grows
with population sizes. In all cases the simulations were ran for at least twice the time to
speciation of the DH model, Fig. \ref{time}(a) to make sure the similarity distribution would either
move to the left of $q_{min}$ and speciate (for large enough $B$) or stay frozen close to
$q=q_{min}$. Error bars for  $B$ are of the order of symbols size:  $\pm 10^4$ for $B \geq 10^5$ and
$\pm 5 \times 10^3$ for $B < 10^5$.

\section{Spatial model with finite genomes}
\label{SM}

The importance of space in evolution has long been recognized
\cite{Wright-1943,Rosen-1995,coyne_speciation_2004,nostil_2012} and explicit empirical evidence of its role has
been recently provided by ring species
\cite{irwin_speciation_2005,martins_evolution_2013,Martins-2016}. The spatial model we discuss here
is a simplified version of the model proposed in \cite{de_aguiar_global_2009}. The main additional
ingredient with respect to the DH model with finite genomes is that the individuals are now
distributed on a two-dimensional $L\times L$ square area with periodic boundary conditions. Mating
is not only restricted by genetic similarity but also by spatial proximity, so that an individual
can only choose as mating partner those inside a circular neighborhood of radius $S$ centered on its
spatial location, called the {\it mating neighborhood}. We note that a number of other effects, such
as demographic stochasticity \cite{mckane2016}, population expansions
\cite{martins_evolution_2013,Goodsman-2014}, costs of reproduction \cite{lecunff-2014}, and
migration rates between subpopulations \cite{Yamaguchi-2013} might also influence the outcome of
speciation. Here we consider only the effect of finite mating neighborhoods and keep all the other
ingredients as similar as possible to the original DH model.

\begin{figure}
	\centering \includegraphics[scale=0.35]{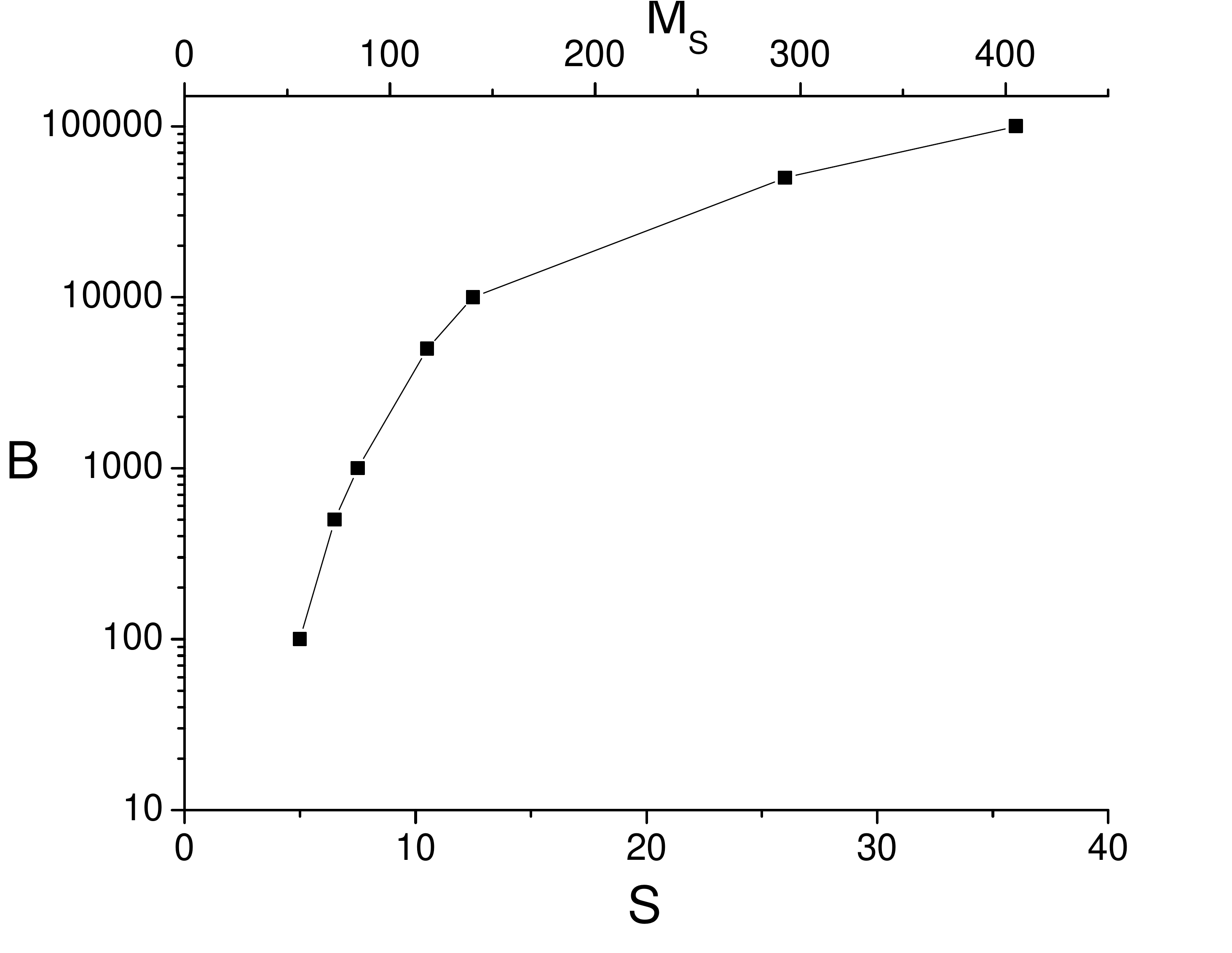}
	\caption{Minimum number of genes required for speciation as a function of the mating neighborhood 
		radius $S$. The top axis show the corresponding average number of individuals
		in the mating neighborhood ($M=1000$, $\mu=1/4000$.)} 
	\label{fig3}
\end{figure}

Space represents an environment where resources are distributed homogeneously and the individuals
should occupy it more or less uniformily so that enough resources would be available to all of them.
The total carrying capacity is the population size $M$ and the average area needed per individual is
$M/L^2$. The dynamics is constructed in such a way that offspring are placed close to the location
of the original parents and the approximately  uniform distribution of the population is preserved
at all times. However, the mechanism used in the DH model of picking a random individual from the
population to be the first parent and then a second individual to be the second parent (and
repeating the process $M$ times) promotes strong spatial clustering.

\begin{figure}
	\centering 
	\includegraphics[scale=0.27]{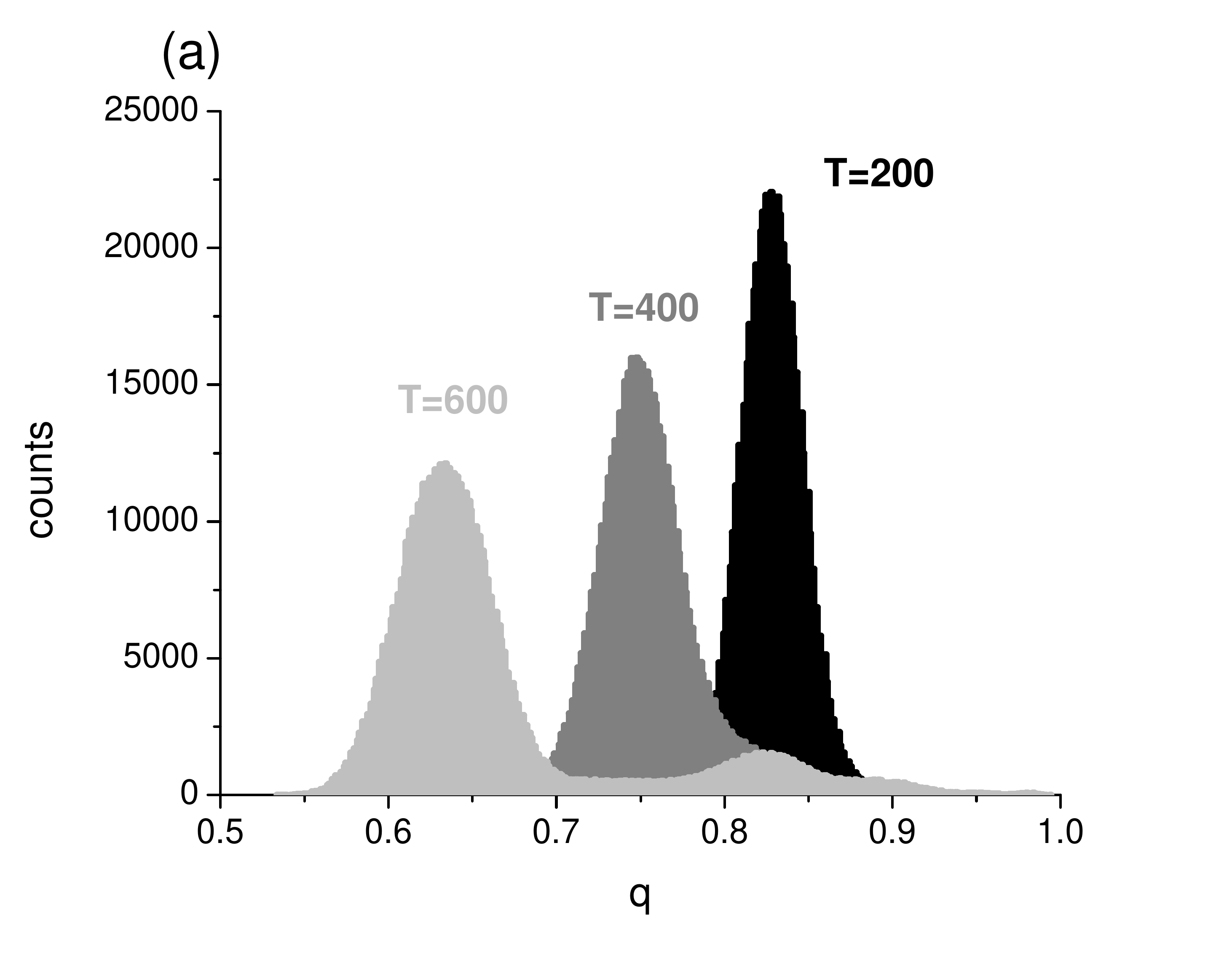} \includegraphics[scale=0.27]{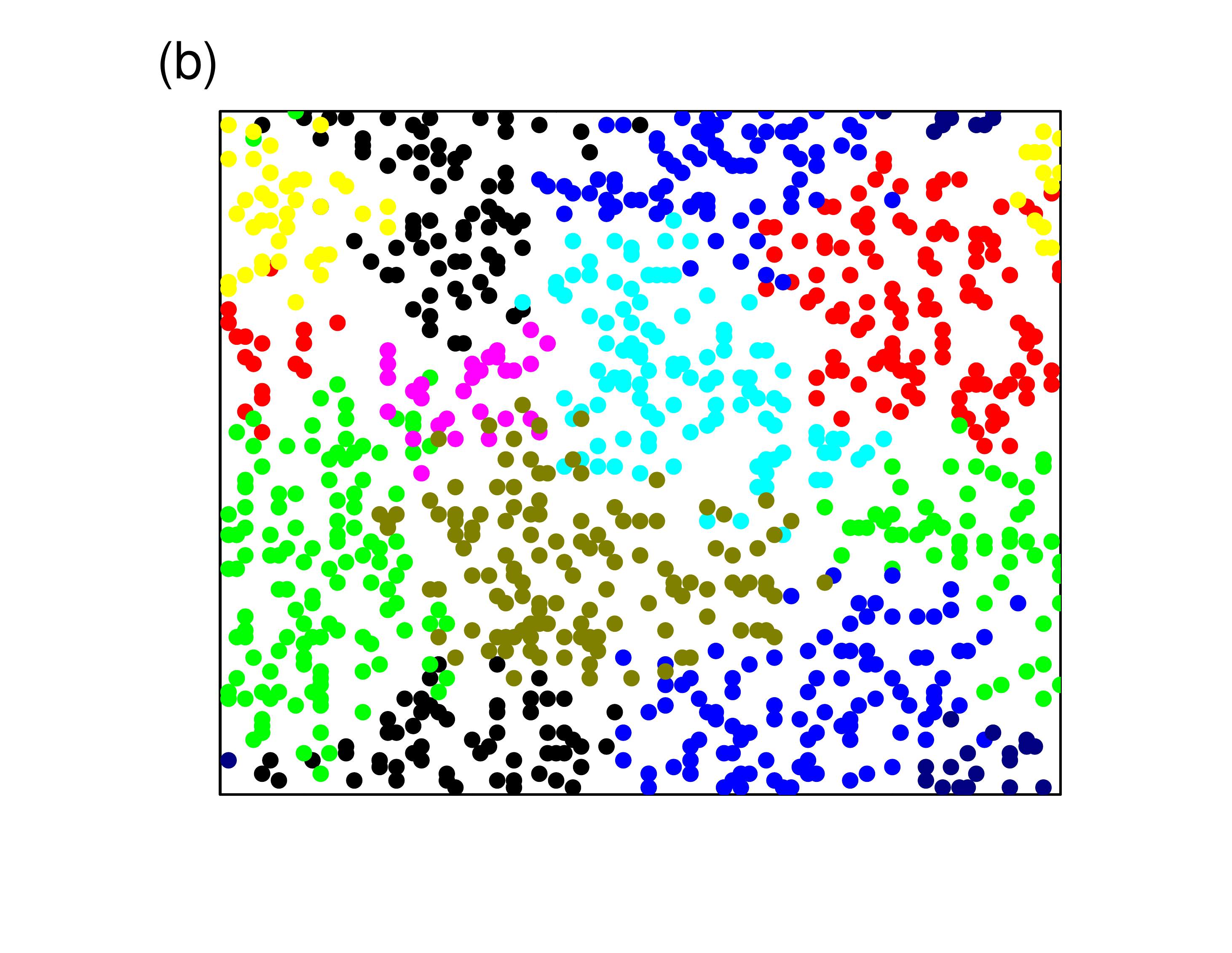}
	\includegraphics[scale=0.27]{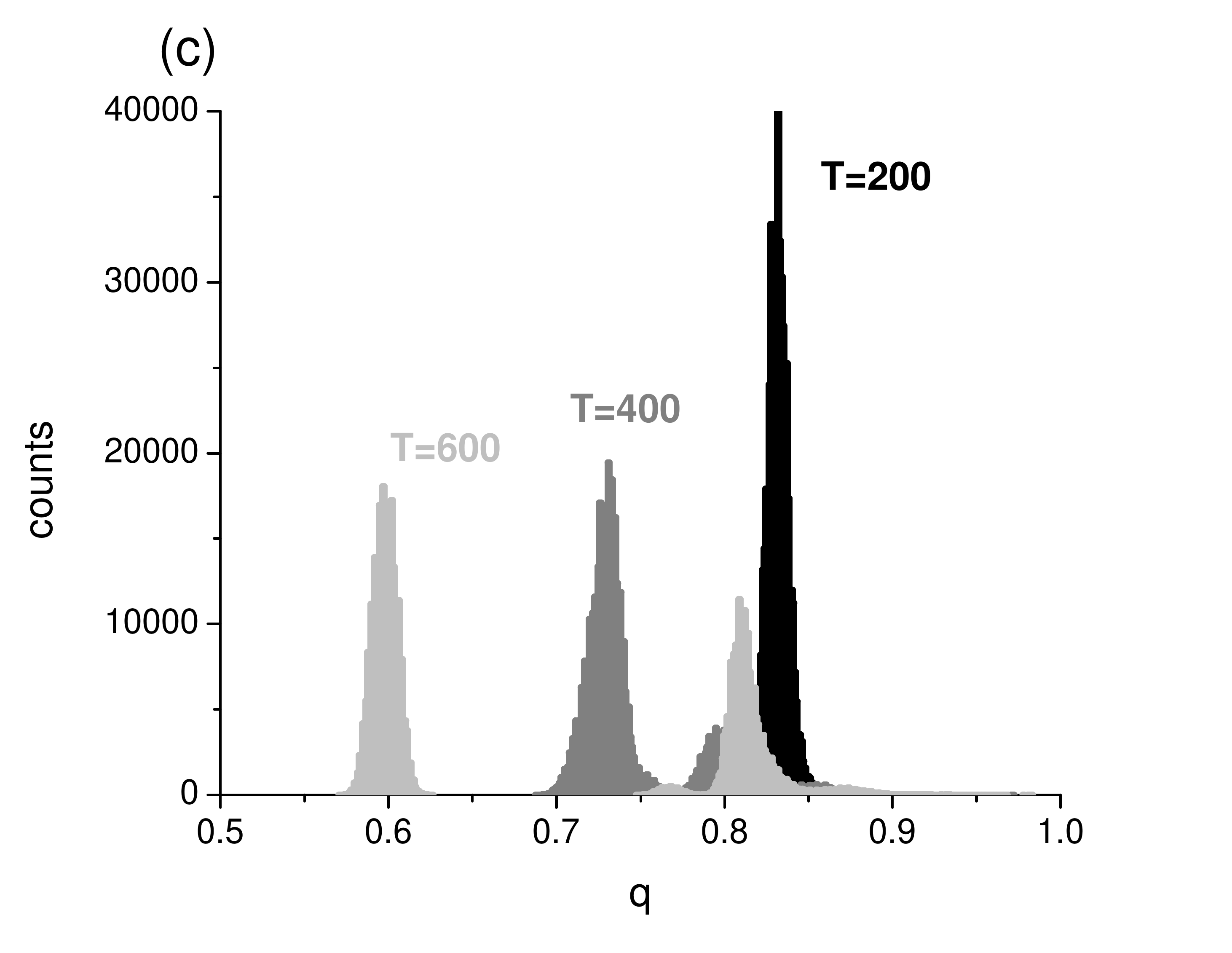} \includegraphics[scale=0.27]{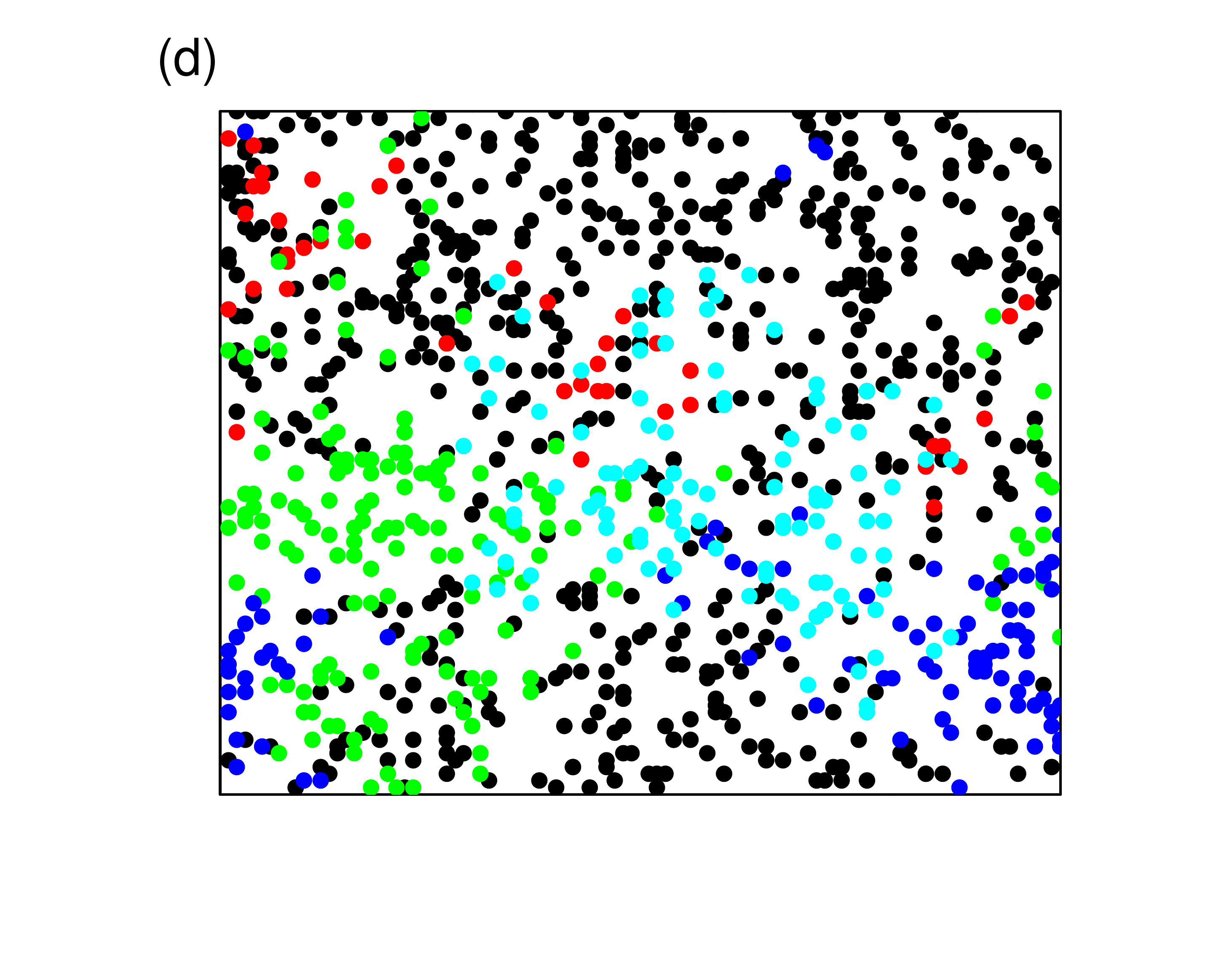}
	\includegraphics[scale=0.27]{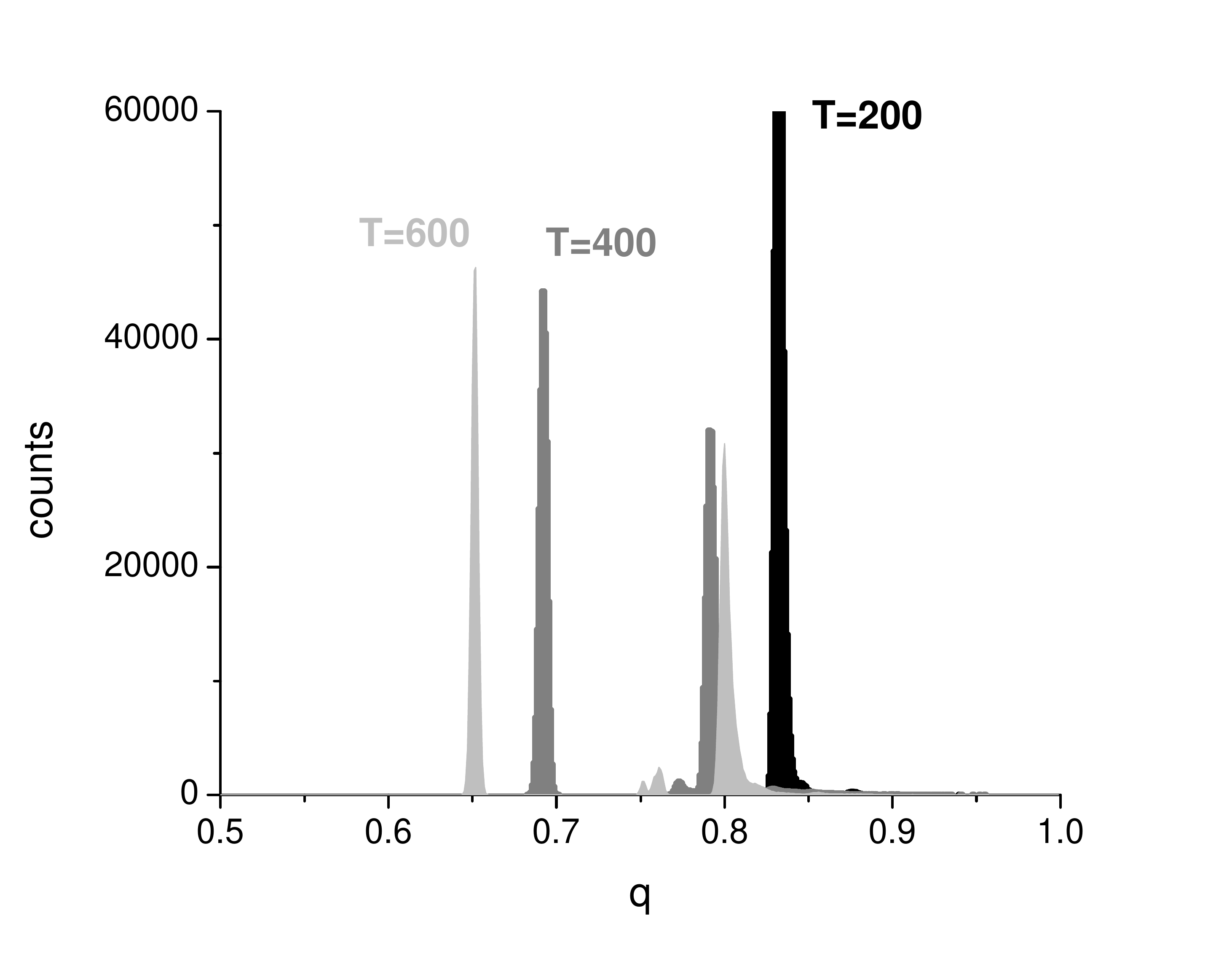} \includegraphics[scale=0.27]{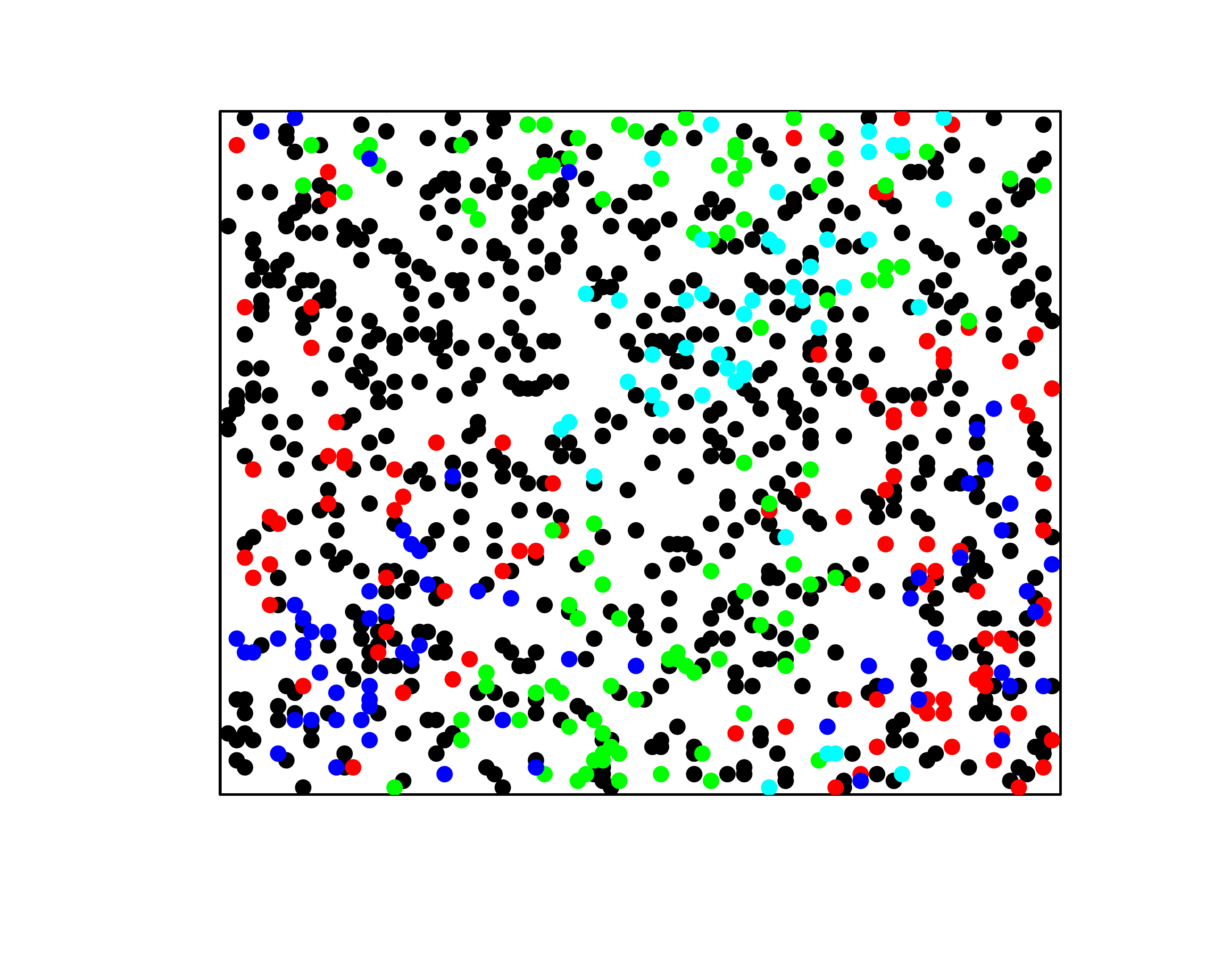} 
	\caption{(color online) (a) Distribution of similarity coefficients for the spatial model with $B=1000$,
		$q_{min}=0.8$ and $S=7$ for $T=200$, $400$ and $600$ generations. (b) Spatial distribution of the
		population at $T=600$; different colors (shades of gray) show different species. Panels (c) and (d) show similar
		plots for $B=10000$ and $S=11$; (e) and (f) show the results for $B=100000$ and $S=14$.}
	\label{fighds}
\end{figure}

In order to avoid clustering we implement the dynamics in a slightly different way
\cite{de_aguiar_global_2009} (see also section \ref{VLG}): the initial population is randomly placed
in the $L \times L$ area. Each one of the $M$ individuals has a chance of reproducing but there is a
probability $Q$ that it will not do so, accounting for the fact that not all individuals in the
present generation will be first parents of the next. In the case the {\it focal} individual does
not reproduce, another one from its mating neighborhood is randomly chosen to reproduce in its
place.  In either case the offspring generated will be positioned exactly at the location of the
focal individual or will disperse with probability $D=0.01$ to one of 20 neighboring sites.
Therefore, close to the location of every individual of the previous generation there will be an
individual in the present generation, keeping the spatial distribution uniform and avoiding the
formation of clusters. The first parent (or a neighbor reproducing in its place) chooses a compatible
second parent within its mating neighborhood of radius $S$. If $S$ is small, the number of
individuals in the mating neighborhood can be close to zero due to fluctuations in the spatial
distribution. To avoid this situation, and also to follow the procedure introduced in
\cite{de_aguiar_global_2009}, if the number of compatible mates in the neighborhood is smaller than
$P$ ($P=3$ in the simulations), the individual expands the search radius to $S+1$. If the number of
compatible mates is still smaller than $P$, the process is repeated twice more up to $S + 3$, and if
there is still less than $P$ potential mates another neighbor is randomly selected to reproduce in
its place \cite{de_aguiar_global_2009}. The probability $Q$ was fixed in $Q=e^{-1} \approx 0.37$,
which corresponds approximately to the probability that an individual is not selected in $M$ trials
with replacement, $(1-1/M)^M \approx e^{-1}$, in accordance with the DH model.

The size of mating neighborhood is a key extra parameter. If $S \simeq L$ we recover the DH model
with finite genomes. If $S$ is small speciation is strongly facilitated and can occur for much
smaller values of $B$ for a given $q_{min}$ or $G/B$. Fig. \ref{fig3} shows the minimum value of $B$
for which speciation happens as a function of the size of the mating neighborhood $S$ for
$q_{min}=0.8$. The figure also shows the average number of individuals inside the mating
neighborhood, $M_S = M \pi S^2/L^2$. The values were obtained by varying $S$ by units of $0.5$ when
$B \leq1000$ and by units of $2$ for larger $B$'s. Populations evolved for  $T=600$ generations. In
some cases speciation did occur for slightly smaller values of $B$, but it took much longer times.
For $B=1000$, for instance, we observed speciation for $S=8$ for $T \approx 3000$. Comparing with
the finite DH model we note that speciation can take place even for $B=100$ if $S \leq 5$. As $B$
increases the restriction in the size of the mating neighborhood required for speciation decreases,
until it is unnecessary if $B$ is of the order of 110000.

Figure \ref{fighds} shows the histogram of similarity between pairs of individuals at three times and a
snapshot of the population for $B=1000$, $S=7$; $B=10000$, $S=11$ and $B=100000$ and $S=14$. The
peaks in the distribution are larger both because $B$ is small and because of the restriction in
$S$. For $B=1000$ the species are well localized in space, with little overlap at their boundaries
(see \cite{de_aguiar_global_2009}). For $B=10000$ and $B=100000$ the spatial overlap between species
is considerably larger.

\section{Spatial model  with very large genomes}
\label{VLG}

As discussed in section \ref{SM}, the construction of spatial models that converge to the finite DH model
as the parameter S becomes large is not straightforward. The problem resides in the way generations are
constructed in the DH model, where a first parent is chosen at random from the population and then a
second, genetically compatible, parent is also chosen at random to generate an offspring. The direct
application of this procedure in the spatial model would consist in choosing a random first parent
and picking the second parent from its mating neighborhood, an area $\pi S^2$ centered on the
individual. The offspring should be put close to the location of the first or second parent, or
somewhere in between. This, however, leads to spatial clustering of the population, since a large
fraction ($\approx e^{-1}$) of the population is never chosen as first parents, leaving holes in these
areas and overcrowding areas where individuals are picked twice or more. To avoid this situation we
have replaced the random choice of the first parent by going through the population one by one and
giving each individual a chance ($1-e^{-1}$) to reproduce. When it did not we picked another individual
from its mating neighborhood to reproduce in its place, instead of a random individual taken from
the entire population as in the original DH or finite DH models. The offspring is always placed
close to the location of the first parent, keeping the distribution spatially uniform. 

\begin{figure}
	\centering 
	\includegraphics[scale=0.27]{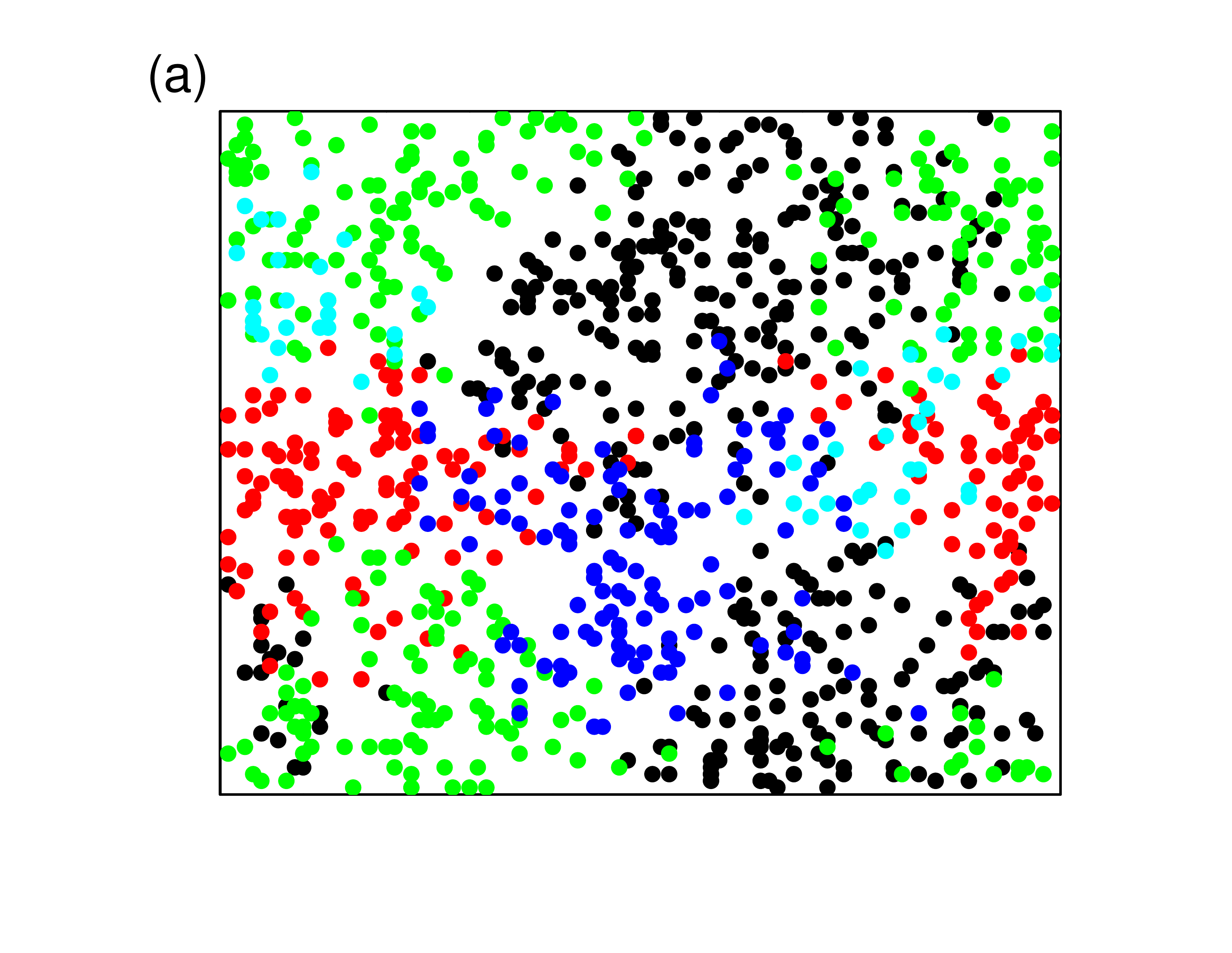} 
	\includegraphics[scale=0.27]{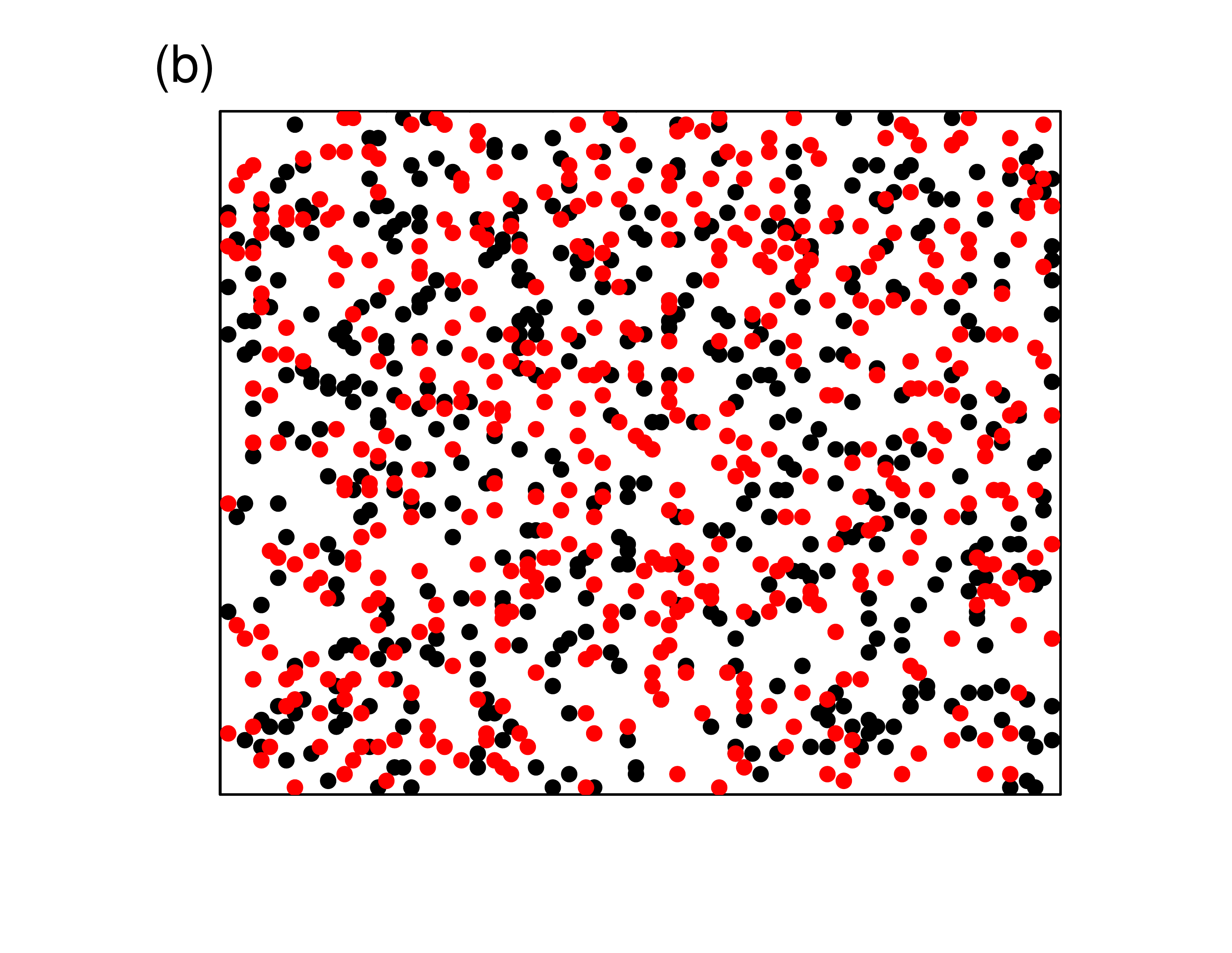}
	\caption{(color online) Spatial distribution of the population for $B=200000$ at $T=300$ for 
		(a) $S=10$ and (b) $S=40$. Small mating neighborhoods favors spatial isolation of species.}
	\label{fig5}
\end{figure}

In order to compare the spatial model with an equivalent sympatric system we implemented a variation
of the finite DH model in which, similarly to the spatial model, we give each individual a chance
($1-e^{-1}$) to reproduce and, when it does not, we pick another random individual from the entire
population to reproduce in its place. The results obtained with this variation are qualitatively
identical to those described in sections \ref{DH} and \ref{DHF}. This validates the comparison of
the DH finite genome model with the spatial model introduced in section \ref{SM}.

Finally we consider the process of speciation in the spatial model for genome sizes $B$ above the
threshold where speciation occurs in the sympatric model. In this case speciation occurs for any
value of $S$, small or large. For small $S$ local mating creates a spatial distribution of genotypes
where nearby individuals tend to be similar but different from others located far away
\cite{de_aguiar_global_2009}. For finite populations this genetic gradient is not smooth, but
step-like, and prone to break up into genetically isolated groups that are spatially correlated.
Speciation happens not because $B$ is large, but because $S$ is small. For large $S$, on the other
hand, the mechanism is very different and takes place on the global scale of the population. The
balance between mutations, which generally increases the average genetic distance between two
individuals, and sexual reproduction, which mixes the genomes and has the opposite effect, leads to
a distribution of genetic distances in the population. Only when this distribution becomes wide
enough, as compared with the criterion of assortative mating,  the population splits into species. 
Figure \ref{fig5} shows the resulting populations in each case, where a signature of $S$ is clearly
seen in the spatial organization of the species.

\section{Discussion}

The mechanisms responsible for the origin of species remain controversial \cite{Chung_2004}. Among
the important questions yet to be fully understood is the role of geography (allopatric, parapatric and
sympatric modes) and the number of genes involved in the evolution of reproductive isolation. It has been
recently argued that these two points are closely related and that detailed molecular analysis might
reveal the geographic mode behind specific speciation events \cite{Machado_2002}.
 
Sympatric speciation has been thought to be the most unlikely of the modes, since the possibility of
constant gene flow would keep the populations mixed and prevent the evolution of reproductive
barriers \cite{Bolnick_2007}. Dieckmann and collaborators have argued that this is not so if
competition for resources is strong enough \cite{Dieckmann_1999,Dieckmann_2003,Doebeli_2005}. In
that case reproductive isolation would not only evolve naturally but would {\it require} sympatry,
otherwise competition for local resources would not be intense \cite{Dieckmann_1999}.

Surprisingly, Derrida and Higgs have shown that sympatric speciation may occur even without
competition, in a totally neutral scenario, if mating is assortative and if the number of loci in
the genome can be considered infinite \cite{higgs_stochastic_1991}. When loci are interpreted as
genes the hypothesis of an infinitely large genome becomes rather unrealistic. However, at the
molecular level, where nucleotide sequences are the units to be considered, infinite loci models
become attractive \cite{Kimura_1964,Ewens_1979}. Nevertheless, nucleotides can hardly be considered
independent and certainly do not segregate in the way as assumed by the DH models. Understanding how
many independently segregating loci are necessary for neutral sympatric speciation is, therefore, an
important question. The main parameters controlling the dynamics in the DH model are $q_0 = (1+ 4\mu
M)^{-1}$ and the assortativity measure $q_{min}$.  The combination $\theta = 2 \mu M$ also appears
in Hubbell's neutral theory of biodiversity \cite{Hubbell-2001} and is the {\it fundamental
	biodiversity number}, since it controls the number of species in a community and the abundance
distribution.

In this paper we have revisited the DH model and simulated it for finite numbers of loci. We found
that, for typical parameters used in the original paper, speciation happens only for very large
number of loci, of the order of $10^5$. When the number of loci is small, the genetic variability
within the population is large, hindering speciation. The histogram of genetic similarity between
individuals evolves into a broad peak instead of multiple sharp peaks that represent groups of very
similar individuals that are dissimilar among different groups. As the number of genes increase the
peaks become thiner, secondary structures appear and eventually turn into species. Increasing the mutation 
rate and keeping $\mu M$ fixed decreases the minimum number of loci needed and also
the time to speciation.

In contrast with the sympatric DH model, the spatial model introduced in
\cite{de_aguiar_global_2009}  displays speciation with much smaller number of loci. The model
considers a population that is uniformly distributed in space, without any explicit separation into
demes or subpopulations. Although it falls into the class of parapatric models, it has been termed
topopatric to distinguish it from models involving demes or metapopulations
\cite{Manzo_Peliti_1994}. Mating is restricted not only by genetic similarity but also by spatial
proximity, allowing gene flow across the entire population but substantially reducing the speed of
the flow. Mutations are transmitted diffusively across the population, and not instantaneously like
in the sympatric model, largely facilitating speciation \cite{Schneider2016}. Previous studies with
metapopulation models (and infinitely large genomes) observed similar effects, with speciation
occurring for smaller mutation rates due to the isolation of subpopulations
\cite{Manzo_Peliti_1994}. For small number of genes, of the order of 100, speciation occurs only
with severe spatial mating restrictions. Accordingly, the species that form display strong spatial
segregation, with little overlap between adjacent species (Figs. \ref{fighds}(b) and \ref{fig5}(a)).
This type of geographic distribution leads back to question of modes of speciation and suggests that
it might happen that species appeared not because the populations were geographically separated (as
in allopatry) but rather they are geographically separated because they emerged in a homogeneous
environment with slow gene flow. If the number of loci participating in the assortative process is
large the spatial restriction on mating can be relaxed. As a consequence, the spatial segregation
decreases and the overlap among species increases.\\ \\

\noindent Acknowledgments: 

\noindent It is a pleasure to thank David M. Schneider, Ayana Martins and Blake Stacey for their
critical reading of this paper and many suggestions. This work was partly supported by the Brazilian
agencies FAPESP (grant 2016/06054-3) and CNPq (grant 302049/2015-0).

\clearpage

\end{document}